\documentclass[%
reprint,
superscriptaddress,
 prl,
 amsmath,
 amssymb,
 aps,
 longbibliography
]{revtex4-2}

\usepackage{cancel}
\usepackage{graphicx}
\usepackage{dcolumn}
\usepackage{bm}
\usepackage[usenames,dvipsnames]{color}
\usepackage{xcolor}
\usepackage{soul}
\usepackage{subcaption}
\usepackage{enumitem}
\usepackage{xspace}
\usepackage{booktabs}
\usepackage{lipsum}
\usepackage{siunitx} 
\usepackage[ISO]{diffcoeff}
\usepackage{subcaption}
\usepackage{ragged2e}
\usepackage{mathrsfs}
\usepackage{hyperref}
\usepackage[capitalise]{cleveref}
\usepackage[normalem]{ulem}
\usepackage{amsmath}
\usepackage{wrapfig}
\usepackage{esvect} 
\usepackage{titlesec}
\usepackage{titletoc}
\usepackage{chngcntr}
\usepackage{fontawesome5}
\usepackage{mleftright}
\usepackage[nodayofweek]{datetime}
\usepackage{comment}
\usepackage{multirow}
\renewcommand{\v}[1]{\boldsymbol{#1}}	

\DeclareCaptionJustification{justified}{\justifying}
\definecolor{papercolor}{HTML}{b22222}
\hypersetup{
    colorlinks=true,       
    linkcolor=papercolor,          
    citecolor=papercolor,        
    filecolor=papercolor,      
    urlcolor=papercolor           
}

\definecolor{orcid-green}{RGB} {166, 206, 57}
\newcommand{\MYhref}[3][blue]{\href{#2}{\color{#1}{#3}}}%

\titleclass{\mysection}{straight}[\section]
\titleformat{\mysection}[runin]
  {\itshape}{\thesection}{}{}[.---]
\titlespacing{\mysection}{1em}{1em}{0em}

\newcommand{\dd}{\mathop{}\!\mathrm{d}}
\newcommand{\dv}[2]{\frac{\mathrm{d} #1}{\mathrm{d} #2}}
\newcommand{\pv}[2]{\frac{\partial #1}{\partial #2}}
\newcommand{\pvl}[2]{\partial #1/\partial {#2}}

\newcommand{\mrm}[1]{\mathrm{#1}}

\newcommand{\bv}[1]{\mathbf{#1}}
\newcommand{\bs}[1]{\boldsymbol{#1}}
\newcommand{\hbv}[1]{\hat{\mathbf{#1}}}
\newcommand{\mcal}[1]{\mathcal{#1}}

\begin{document}

\title{\boldmath A Levitated-Magnet Vector Force Sensor for Spin-Dependent Exotic Interactions}

\author{Dorian W.~P.~Amaral\,\MYhref[orcid-green]{https://orcid.org/0000-0002-1414-932X}{\faOrcid}}
\email{damaral@ifae.es}
\affiliation{Institut de F\`{ı}sica d’Altes Energies (IFAE), The Barcelona Institute of Science and Technology,
Campus UAB, 08193 Bellaterra (Barcelona), Spain}

\author{Lei Cong\,\MYhref[orcid-green]{https://orcid.org/0000-0003-0002-1840}{\faOrcid}}
\email{conglei1@uni-mainz.de}
\affiliation{Helmholtz Institut Mainz, 55099 Mainz, Germany}
\affiliation{GSI Helmholtzzentrum f\"{u}r Schwerionenforschung GmbH, 64291 Darmstadt, Germany}
\affiliation{Institut f\"{u}r Physik, Johannes Gutenberg-Universit\"{a}t Mainz, 55099 Mainz, Germany}

\author{Tim M. Fuchs\,\MYhref[orcid-green]{https://orcid.org/0000-0001-6367-3948}{\faOrcid}}
\affiliation{School of Physics and Astronomy, University of Southampton,
SO17 1BJ, Southampton, UK}

\author{Hendrik Ulbricht \,\MYhref[orcid-green]{https://orcid.org/0000-0003-0356-0065}
{\faOrcid}}
\affiliation{School of Physics and Astronomy, University of Southampton,
SO17 1BJ, Southampton, UK}

\author{Dmitry Budker\,\MYhref[orcid-green]{https://orcid.org/0000-0002-7356-4814}
{\faOrcid}}
\affiliation{Helmholtz Institut Mainz, 55099 Mainz, Germany}
\affiliation{GSI Helmholtzzentrum f\"{u}r Schwerionenforschung GmbH, 64291 Darmstadt, Germany}
\affiliation{Institut f\"{u}r Physik, Johannes Gutenberg-Universit\"{a}t Mainz, 55099 Mainz, Germany}
\affiliation{Department of Physics, University of California at Berkeley, Berkeley, California 94720-7300, USA}

\begin{abstract}
\noindent
We present a magnetically levitated ferromagnetic vector force sensor that enables selective searches for spin-dependent exotic interactions mediated by beyond-Standard-Model bosons. A defining feature of spin-dependent exotic interactions is that they can generate forces with distinct directional signatures set by the relative spin configuration of the interacting bodies. We show that our sensor resolves these signatures by mapping forces along different axes onto distinct translational modes with different resonance frequencies, thereby separating interaction channels within the same coupling class. As a representative example, we study parity-violating axial-vector--vector interactions mediated by a spin-1 $Z'$ boson between a sensing and a driving levitated ferromagnet. Using a matched-filter likelihood analysis, we show that a 
setup based on an already demonstrated experiment
can probe the pure electron--electron coupling $g_A^e g_V^e$ in the previously inaccessible force range $\lambda \lesssim 1\,\mathrm{cm}$, corresponding to mediator masses $M_{Z'} \gtrsim 10^{-5}\,\mathrm{eV}/c^2$.
Our results establish levitated ferromagnets as a promising platform for millimeter-scale searches for spin-dependent fifth forces and for resolving the multiple effective potentials associated with a given coupling class.
\end{abstract}

\maketitle

\mysection{Introduction}

Searching for spin-dependent exotic interactions provides a sensitive low-energy window into physics beyond the Standard Model \cite{cong_spin-dependent_2025,safronova_search_2018}. Such interactions can be mediated by light bosons from many well-motivated extensions of the Standard Model \cite{dine_origin_2003,feng_dark_2010,graham_cosmological_2015,navas_review_2024}, leading to exotic forces between spin-polarized and unpolarized matter \cite{moody_new_1984,dobrescu_spin-dependent_2006}. A broad range of precision-measurement platforms has been developed to probe these interactions \cite{cong_spin-dependent_2025,gaul_constraints_2026,shu_constraint_2024,jiao_experimental_2021,tian_constraints_2025,ji_new_2018,kim_experimental_2019,heckel_preferred-frame_2008}, including atomic and molecular parity-violation experiments, atomic interferometers, nitrogen-vacancy centers, comagnetometers, and macroscopic force-sensing experiments.

A central challenge in searches for spin-dependent exotic interactions is selective discrimination---the ability to distinguish among fermion-pair channels and effective potentials. In simple atomic systems, where the constituent particles are few and well defined, individual fermion-pair contributions may be treated explicitly \cite{karshenboim_precision_2010,cong_improved_2025}. In macroscopic experiments, however, the source and sensor contain ensembles of different fermion species, so that multiple fermion-pair channels contribute simultaneously \cite{yan_new_2013}. The modern theoretical framework further makes explicit that, for a given coupling type, a single mediator can generate several effective potentials \cite{cong_spin-dependent_2025,dobrescu_spin-dependent_2006,fadeev_revisiting_2019,cong_improved_2025,kang_exotic-interaction_2025}. For example, a $Z’$ boson with parity-violating axial-vector–vector (AV) couplings induces both $V_{12\pm13}$ and $V_{11}$~\cite{cong_spin-dependent_2025}. The observable signal can thus involve multiple combinations of fermion-pair channels and effective potentials. Going beyond the common simplification that certain potential terms or fermion-pair couplings can be neglected \cite{kim_experimental_2019,su_search_2021}, achieving selective discrimination has emerged as one of the key problems in searches for spin-dependent exotic interactions.

\begin{figure*}
    \begin{center}
    \includegraphics{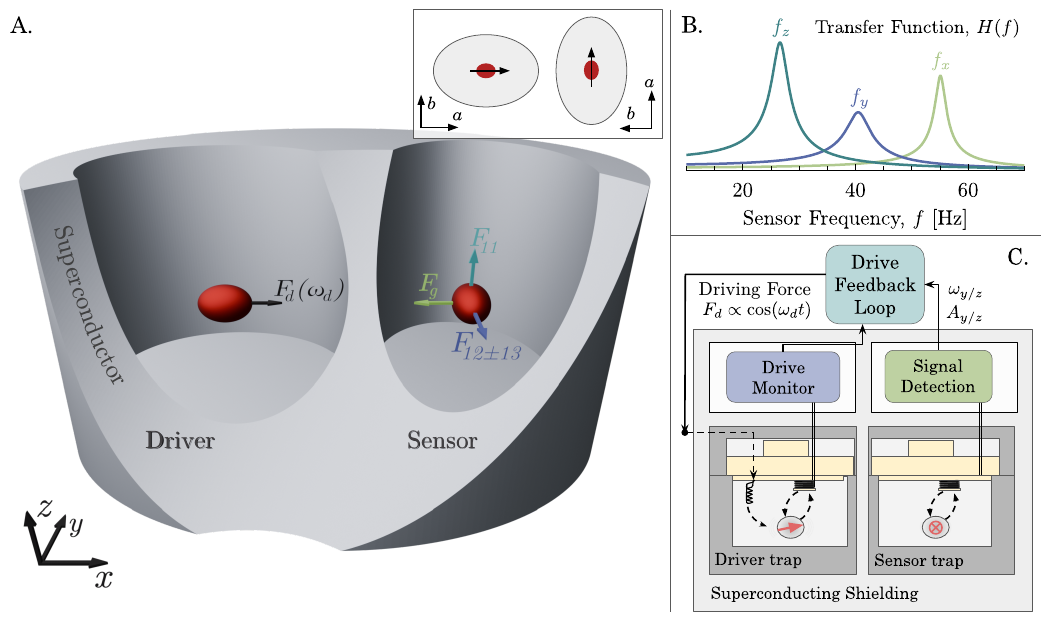}
    \end{center}
    \vspace{-3ex}
    \caption{
    Depiction of the experiment. The time-altering potentials sourced by the driver generate a force on the sensor particle, amplifying its motion when the drive frequency matches a mode frequency. \textbf{A.} A sliced 3D render of the setup. Motion of the driver along the $x$-axis  generates the forces $\bv{F}_{11}$ and $\bv{F}_{12\pm13}$ on the sensor. Inset shows a top-down view of the traps with the semi-major ($a$) and semi-minor ($b$) axes of the traps and the particle magnetic moments. \textbf{B.} An example force-displacement transfer function $H(f)$ of the sensing trap. \textbf{C.} Schematic implementation of the detection, trapping, shielding and feedback in the experiment. Our results are agnostic to the exact implementation.}\label{Fig:Schematic}
\end{figure*}

Magnetic levitation has become a powerful platform for realizing ultralow-dissipation mechanical systems by eliminating mechanical contact, thereby suppressing clamping losses and enabling exceptionally high mechanical quality factors and long coherence times
\cite{cirio_quantum_2012,romero-isart_quantum_2012,gonzalez-ballestero_levitodynamics_2021,timberlake_acceleration_2019,hofer_high-q_2023,gutierrez_latorre_superconducting_2023,ji_levitated_2025,vinante_ultralow_2020}.
Depending on the response of the levitated object, distinct levitation mechanisms have been realized, including ferromagnetic \cite{fuchs_2024_gravity,peng_ferromagnetic_2026}, diamagnetic \cite{ji_levitated_2025,tian_constraints_2025}, and superconducting platforms \cite{carney_superconducting_2025,higgins_maglev_2024}. 
Among these, superconducting levitated ferromagnetic oscillators are particularly attractive for precision sensing ($0.5~\mrm{fN\,Hz^{-1/2}}$~\cite{fuchs_2024_gravity}) because of their large mass, tunable trap frequencies, and compatibility with ultrasensitive magnetic readout techniques such as SQUID detection \cite{timberlake_acceleration_2019,vinante_surpassing_2021,sheng_levitated_2026}. 
These features have already motivated applications to searches for new physics, including long-range non-Newtonian forces \cite{amaral_magnetic_2026}, ultralight dark matter \cite{amaral_first_2025}, and ultraheavy dark matter~\cite{amaral_towards_2025,qin_mechanical_2025}.

In this work, we exploit these features to develop magnetic levitation into a highly selective, precision vector probe for spin-dependent fifth forces at short force range. Two levitated ferromagnets serve as the driving spin source and the spin sensor, offering three distinct advantages. 
First, they naturally carry macroscopic net electron-spin polarization \cite{vinante_ultralow_2020,wang_dynamics_2019} and are among the highest-spin-density materials \cite{cong_spin-dependent_2025}, enabling sensitivity to effects such as axion-induced torques \cite{li_search_2026,yang_ultralight_2026,fadeev_gravity_2021,kalia_ultralight_2024,ferreira_ultra-light_2021}.
Second, because forces along different spatial directions couple to distinct mechanical modes at different resonance frequencies, the system can be deliberately designed to function as a vector force sensor. 
Third, the small size of the levitated magnets and the integrated superconducting levitation-and-shielding architecture allows exceptionally small source–sensor separations. 
We study parity-violating AV interactions mediated by a spin-1 boson as a concrete example, such as a $Z’$ \cite{gaul_constraints_2026,antypas_isotopic_2019}. While, parity-violating effects have traditionally been explored through small energy shifts and transition amplitudes in atomic and molecular systems \cite{antypas_isotopic_2019,safronova_search_2018,dzuba_long-range_2022}, as well as in the landmark $^{60}$Co beta-decay experiment of Wu \emph{et al.} \cite{wu_experimental_1957}, here we show that parity violation can be probed directly as a macroscopic force signal. Our setup opens a previously inaccessible parameter regime for parity-violating AV electron–electron exotic interactions \cite{heckel_limits_2013,hunter_using_2013}. More broadly, our results point to a new design paradigm for fifth-force experiments based on selective resolution.

\mysection{Directional Structure of Spin-Dependent Exotic Forces}\label{Sec:force_direction}

Spin-dependent exotic interactions differ from spin-independent ones by their richer vector structure, determined by the relative orientations of spin, displacement, and velocity. In our setup, illustrated in \cref{Fig:Schematic}, the parity-violating AV interaction between the source and sensor gives rise to the potentials \cite{cong_spin-dependent_2025}:
 \begin{align}
V_{AV}(\bv{r}, \bv{v}) &= \underbrace{- g_A^e g_V^e \frac{\hbar^2}{4\pi m_e}\big[(\bs{\sigma}_e \times \bs{\sigma}_e') \cdot \hbv{r}\big]\left(\frac{1}{r^2} + \frac{1}{\lambda r}\right) e^{-r/\lambda} }_{ V_{11}|_{AV} + V_{11}|_{VA}} \nonumber\\ 
 &+\underbrace{g_A^eg_V^{e+N} \frac{\hbar}{4\pi}\big[(\bs{\sigma}_e - \bs{\sigma}_e')\cdot\bv{v}\big]\frac{1}{r}e^{-r/\lambda} }_{V_{12+13}|_{AV} + V_{12-13}|_{AV}}\,. 
 \label{eq:potential}
\end{align}
where $\hbar$ is the reduced Planck constant, $\boldsymbol{\sigma}_{e}$ and $\v\sigma_e^{\,\prime}$ are vectors of Pauli matrices representing the spins $\boldsymbol{s}_i=\hbar \boldsymbol{\sigma}_i/2$ of the two electrons, $m_e$ is the electron mass, $\lambda$ is the force range (which is related to the mass of a new $Z'$ via $M_{Z'}=\hbar/(c\lambda)$, with $c$ the speed of light), and $r$ is the distance between the two fermions.
Here, $g_V^{e+N}\equiv g_V^e+g_V^N$ is defined to represent the case in which the exotic interaction at the vector vertex involves electrons and nucleons for $V_{12\pm13}$. Note that we study pairs of potentials here; see Ref.\,\cite{cong_spin-dependent_2025} for further discussion. 

The exotic potentials in Eq.~(\ref{eq:potential}) generate the corresponding exotic forces $\bv{F}_{11}(\bv{r})$ and $\bv{F}_{12\pm13}(\bv{r},\bv{v})$, which can be derived from the Euler--Lagrange equations~\cite{supp_mat}. These forces act together with the gravitational interaction between the magnets, $\bv{F}_g(\bv{r})$, such that the total force acting on the sensing magnet is
$\bv{F}(\bv{r},\bv{v}) \equiv \bv{F}_g(\bv{r}) + \bv{F}_{12\pm13}(\bv{r}, \bv{v}) + \bv{F}_{11}(\bv{r})$. A schematic of the driver--sensor configuration is shown in Fig.~\ref{Fig:Schematic}. Both the source and sensing bodies are levitated magnets, and the source magnet is driven sinusoidally with an amplitude $\mcal{A}$ and known frequency $\omega_d$. The sensing magnet possesses displacement and velocity vectors $\bv{r}$ and $\bv{v}\equiv\dot{\bv{r}}$ relative to the driving magnet. The experiment can be designed such that these forces predominantly excite orthogonal translational modes of the sensing magnet. In particular, by orienting the sensing magnet along $\hbv{s}_e=\hbv{y}$ and configuring the driving magnet such that $\hbv{s}_e'=\hbv{x}$, with displacement along $\hbv{x}$ and driving velocity $\bv{v}(t)=v(t)\,\hbv{x}$, the total force on the sensing magnet is given by
\begin{align}
    &\bv{F}_{g}(\bv{r}) = - \frac{G_N m_sm_d}{r^2}\,\hbv{x}\,,\nonumber\\
    &\bv{F}_{12\pm13}(\bv{r}, \bv{v}) = - g_A^e g_V^{e+N} \mcal{N}_\sigma \mcal{N}_f\frac{\hbar}{4\pi}e^{-r/\lambda} v\left(\frac{1}{r^2} + \frac{1}{\lambda r}\right) \,\hbv{y}\,,\nonumber\\
    &\bv{F}_{11}(\bv{r}) = g_A^e g_V^e \mcal{N}_\sigma^2\frac{\hbar^2}{4\pi m_e}e^{-r/\lambda}\left(\frac{1}{r^3} + \frac{1}{\lambda r^2}\right) \,\hbv{z}\,.
\label{eq:forces}
\end{align} 
Here, $G_N$ is Newton's gravitational constant, $m_{s/d}$ the mass of the sensing/driving magnet, $\mcal{N}_\sigma$ the total number of polarized electron spins in a magnet, and $\mcal{N}_f$ is the total number of fermions.
For simplicity, we assume identical driving and sensing magnets and treat both as point particles, valid for $R \ll \lambda$, where $R$ is the magnet radius. Full expressions for the exotic forces in arbitrary configurations are given in Ref.~\cite{supp_mat}.

\mysection{The  Vector Force Sensor and the Setup}

We develop a spin-polarized source–sensor configuration from the experimentally realized levitated ferromagnetic resonator platform of Ref.~\cite{fuchs_2024_gravity}. In that experiment, a single milligram-scale levitated mass (a Nd$_2$Fe$_{14}$B permanent magnet) was used to detect short-range gravitational forces \cite{vinante_ultralow_2020,fuchs_2024_gravity}. Here, instead, we exploit the macroscopic spin of such a levitated body and design the sensing magnet to act as a vector force sensor, thereby enabling selective resolution. As an initial demonstration design, we consider a double-trap setup, permitting better alignment and suppressing several noise sources. 

Matching to the parameter values of Ref.~\cite{fuchs_2024_gravity}, we use a sensing magnet of mass $m_s \approx 0.356\,\mrm{mg}$, radius $R_s = 236\,\mrm{\mu m}$, and with resonance frequencies $\{f_x, f_y, f_z\} =\{55.1\,\mrm{Hz}, 40.6\,\mrm{Hz}, 26.7\,\mrm{Hz}\} $ along each of the translational modes, where each mode has quality factor $\{Q_x, Q_y, Q_z\} = \{6\times10^{6}, 2\times10^6, 1\times10^7\}$. The traps have semi-major and semi-minor axes of $a = 2.25\,\mrm{mm}$ and $b = 1.75\,\mrm{mm}$, respectively. This setup provides intrinsically spin-polarized source and sensing masses with a macroscopic polarized-electron number, $\mcal{N}_\sigma \simeq 6 \times 10^{18} (m_s / 0.356\,\mrm{mg})$, making it well suited to probes of electron spin-dependent interactions~\cite{supp_mat}.

We drive the motion of the particle in the driver trap sinusoidally along the semi-major axis such that its displacement from its equilibrium position is given by $\bv{r}_d(t) = \mcal{A} \cos(\omega_d t) \hbv{x}$, where $\omega_d$ is the angular frequency of the drive and $\mcal{A}$ is the driving amplitude. The force amplitude can be changed arbitrarily to counteract the shape of the transfer function, negating the need for in-trap tuning of the mode frequency, simplifying the setup, and removing additional noise sources. To mitigate nonlinear effects due to the back-action from the trap walls, we set $\mcal{A} = 0.2 a$, keeping the amplitude to $20\%$ of the trap's semi-major axis. With our proposed configuration in \cref{Fig:Schematic}, the average distance between the sensing and driving particles is then $\langle r \rangle = a + b + \delta$, where $\delta \approx 0.5\,\mrm{mm}$ is the thickness of the wall between the traps. The time-varying force generated by the driver couples to the sensor particle, and we reconstruct the force acting on it from the induced motion.

To discriminate between the orthogonal forces in \cref{eq:forces}, we exploit the distinct resonance frequencies of the trap modes. In mechanical resonators, periodic forces that match the resonance frequencies $\omega_i$, $\omega_d = \omega_i$, lead to an amplification of the motion in the sensor trap along axis $i$. The minimum separation between the mode-frequencies that can be readily distinguished is given by the full-width at half-maximum (FWHM) of the transfer function, which can be expressed in terms of the damping rate of a mode $i$ as $\Delta\omega_i = \gamma_i \equiv \omega_i/Q_i$. To distinguish between modes, we design our experiment such that $\omega_x - \omega_y\geq 2\Delta\omega_y$ and $\omega_{z} - \omega_y \leq 2\Delta\omega_y$, where $\Delta\omega_y$ is the FWHM of the lowest $Q$-factor mode. We show a schematic depiction of this frequency scheme in~\cref{Fig:Schematic}. By designing the the mode frequencies in the sensor trap in this way, a change to the driving frequency shifts the $Q$-factor amplification from one mode in the sensor trap to another, enabling us to selectively amplify one force at a time.

We propose to cool our setup to $T = 4\,\mrm{K}$ in a liquid helium bath, providing a large degree of cooling while enabling a vibrationally quiet environment. As was shown in Ref.\,\cite{fuchs_2024_gravity}, a setup equivalent to the one proposed here has already enabled measurements with an effective mode temperature of $3\,\mrm{K}$, thermally limited at our proposed temperature. The thermal force noise power spectral density (PSD) for the $i^\mrm{th}$ mode then follows from the fluctuation dissipation theorem: 
\begin{align}
\label{eq:thermal_noise}
    &S_{FF,i}^\mrm{th} = 4 m_s k_B T \gamma_i\\
    &\approx 10^{-32}\,\mrm{N^2\,Hz^{-1}}\left(\frac{m_s}{0.4\,\mrm{mg}}\right)\left(\frac{T}{4\,\mrm{K}}\right)\left(\frac{f_i}{35\,\mrm{Hz}}\right)\left(\frac{10^6}{Q_i}\right)\,.\nonumber
    \end{align}
For the measurement to remain thermally limited, stray electromagnetic fields must be kept away from the particles and the magnetic detection circuits. 

Equally, any magnetic coupling between the traps must be eliminated to prevent unwanted crosstalk. Fortunately, the superconducting traps intrinsically provide electromagnetic shielding. Type-I superconductors expel magnetic fields by the Meissner effect, as characterized by the London penetration depth $\lambda_L(T)$. For a lead type-I superconductor, $\lambda_L(\SI{0}{K}) =$ \SI{30.5}{nm} and $\lambda_L(\SI{4}{K}) =$ \SI{160}{nm}~\cite{gasparovic_superconducting_1970}, leading to significant attenuation over even a \SI{5}{\micro m} barrier. Residual leakage can be dominated by defects, openings, and finite-geometry effects. We therefore adopt a conservative design with a minimum shield thickness of $\delta=0.5$ mm between the two traps and seal lengths of at least 2 mm. Additional outer $\mu$-metal shielding can further suppress room-temperature DC fields and reduce flux trapping during cooldown.
A single-trap setup prepared in such a way has shown to be free of noise caused by stray magnetic fields down to a force sensitivity of \SI{30}{aN} within the resonant bandwidth of the mode under test~\cite{fuchs_2024_gravity}. 
The exotic forces will not be attenuated by the superconductor \cite{jackson_kimball_magnetic_2016}.

To assess how the sensitivity can be improved in the near-term, we also propose an optimized setup that employs a larger magnet with radius $R = 1\,\mrm{mm}$ and mass $m_s \approx 31.8\,\mrm{mg}$, as well as a larger trap with dimensions $a = 3.0\,\mrm{mm}$ and $b = 2.3\,\mrm{mm}$.  Larger magnets are advantageous since they carry a greater number of spins and nucleons, while larger traps are preferred if the $Q$-factor is assumed to be Eddy-current limited~\cite{supp_mat}. Our future setup has resonance frequencies $\{f_x, f_y, f_z\} =\{147\,\mrm{Hz}, 104\,\mrm{Hz}, 73.3\,\mrm{Hz}\} $, where each mode has quality factor $\{Q_x, Q_y, Q_z\} = \{4\times10^{6}, 3\times10^6, 3\times10^7\}$. We show how our sensitivities scale with the radius of the magnets and the trap size in \cite{supp_mat}.

\mysection{Analysis}
To derive our projected sensitivities, we use a matched-filter likelihood analysis, which is the optimal linear filter for stationary Gaussian noise \cite{wainstein_extraction_1963}. 
To model the expected Fourier signal, we first consider the forces experienced by the sensing magnet in the time domain. While these forces depend on the relative displacement and velocity of the sensing and driving magnets, the backaction corrections to the trap frequencies and damping rates are negligible since the force gradients in both position and velocity are small~\cite{supp_mat}. The motion of the sensing magnet therefore contributes negligibly to the relative coordinate, allowing us to write $\bv{r}(t) \simeq - \bv{r}_d(t)$ and $\bv{v}(t) \simeq - \dot{\bv{r}}_d(t)$. We then numerically compute our exotic forces via \cref{eq:forces}, where the gravitational force will be unimportant for our analysis since it acts orthogonally to our signal forces. 

Having computed our forces in the time domain, we calculate their discrete Fourier transforms $\hat{F}_{AV}(\omega_k, \lambda)$, where $\omega_k$ is a discrete angular frequency, $AV$ stands for either one of the two  exotic forces of interest, and the templates are dependent on the force range $\lambda$. We focus on the fundamental Fourier components at $\omega=\omega_d$, which carry the dominant signal power and are resonantly enhanced relative to the higher harmonics. For a given total simulation time $T_\mrm{num}$, we normalize these peaks to compute the spectral amplitudes $\tilde{F}_{AV}(\omega_d, \lambda) \equiv \hat{F}_{AV}(\omega_d, 
\lambda) / T_\mrm{num}$; these amplitudes form our signal templates~\cite{supp_mat}. 

To assess the sensitivity of our experiment, we consider a scenario in which we do not observe an exotic signal from additional AV interactions. The only coherent driven motion that will be excited in the sensing magnet will then be due to gravity alone, which acts along the $x$ direction. Therefore, the orthogonal $y$ and $z$ translational modes act as clean signal channels via which we can constrain the exotic forces $F_{12\pm13}$ and $F_{11}$, respectively. We assume an Asimov data set to compute our median expected sensitivities, taking the measured forces in these directions to be zero and accounting for systematics via a pull parameter~\cite{cowan_asymptotic_2011, supp_mat}.

Finally, we derive the projected sensitivity from a one-sided profile-likelihood test statistic. This yields the closed-form expression \cite{supp_mat}: 
\begin{equation}
    \left\lvert g_A^e g_V^{e/e+N}(\lambda)\right\rvert\gtrsim  \frac{\mcal{T}(T_\mrm{obs}/\tau)}{2|\tilde{F}_{AV}^{(1)}(\lambda)|} \sqrt{\frac{q_\mrm{lim} S_{FF}^\mrm{th}(\omega_d)}{T_\mrm{obs}}}\,,
    \label{eq:sensitivity}
\end{equation}
which makes the scaling of the sensitivity transparent. Here, $|\tilde{F}_{AV}^{(1)}|$ is the spectral force amplitude for unit couplings, $\mcal{T}(T_\mrm{obs}/\tau)$ is a dimensionless factor accounting for the finite ring-up of the sensing magnet, and $S_{FF}^\mrm{th}(\omega_d)$ is the thermal force-noise PSD evaluated at the driven mode frequency~\cite{supp_mat}. For our projections, we take $T_\mrm{obs} = 1\,\mrm{day}$, reflecting the total observation time of Ref.~\cite{fuchs_2024_gravity}, and $q_\mrm{lim}\approx 2.71$, corresponding to a $95\%$ confidence level (C.\,L.) projection \cite{supp_mat}.
The relevant product couplings are $g_A^e g_V^e$ for $F_{AV} = F_{11}$ and $g_A^e g_V^{e+N}$ for $F_{AV} = F_{12\pm13}$. The noise PSD is computed from \cref{eq:thermal_noise} for the relevant excited mode. For the uncertainty levels considered here, the systematic correction factor is negligible ($\sigma_\xi\lesssim 50\%$), so that the projected sensitivities are statistics dominated~\cite{supp_mat}.

\mysection{Results and Discussion}

\begin{figure}
    \centering
    \includegraphics{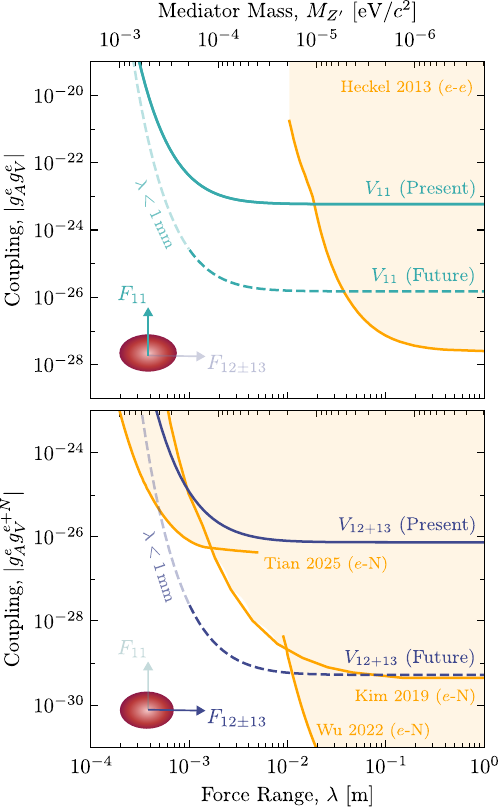}
    \caption{The $95\%$ C.\,L. projected limits on AV interactions. Solid lines represent our sensitivities for an already demonstrated experiment. Dashed lines show what a near-future setup can achieve, turning transparent when the force range becomes smaller than the $1\,\mrm{mm}$ radii of the proposed magnets and the point-particle approximation no longer holds.  \textbf{Top:} Projections for $|g_A^e g_V^e|$. The leading limit on this coupling in this range is from Ref.~\cite{heckel_limits_2013,hunter_using_2013}. \textbf{Bottom:}  Projections for $|g_A^e g_V^{e+N}|$. Current limits are from Refs.~\cite{kim_experimental_2019, wu_experimental_2022,tian_constraints_2025}. }
    \label{fig:sensitivity}
\end{figure}

We develop a vector force sensor for spin-dependent exotic-physics searches. The key enabling feature is that forces with different directional character excite distinct mechanical modes at separate resonance frequencies. This modal selectivity is naturally matched to the directional structure of spin-dependent exotic interactions and enables selective channel discrimination, the central new capability established by our work.

Concretely, we show that a presently realizable levitated-magnet platform can selectively probe AV $e$-$e$ exotic interactions mediated by a spin-1 boson, characterized by $g_A^e g_V^e$ in $V_{11}$, within a previously inaccessible short-range regime of $\lambda \lesssim 10^{-2}\,\mrm{m}$; see \cref{fig:sensitivity}. This relies on bringing the source and sensor to microscopic separations while maintaining macroscopic spin polarization. Compared with torsion-balance spin-source experiments at centimeter scales \cite{heckel_limits_2013} and geoelectron-based searches probing planetary scales \cite{hunter_using_2013}, our levitated-magnet approach is especially suited to unexplored short-range $e$-$e$ spin-dependent interactions. The competitiveness of our setup is also reflected in its projected sensitivity to $g_A^e g_V^{e+N}$: at the level of $10^{-26}$, it is already close to the best result from a recent diamagnetically levitated particle experiment \cite{tian_constraints_2025}. With only modest near-future optimization, the reach could be extended by up to two orders of magnitude in the range $1\,\mrm{mm} \lesssim \lambda \lesssim 1\,\mrm{cm}$.

The platform also enables a direct macroscopic force probe of parity-violating interactions. In contrast to conventional parity-violating experiments for spin-dependent exotic interactions \cite{cong_spin-dependent_2025}, including magnetometer-based searches \cite{kim_experimental_2019,wu_searching_2022,wang_search_2023} and atomic or molecular parity-nonconservation measurements \cite{dzuba_probing_2017,antypas_isotopic_2019}, which infer parity-violating interactions indirectly from small energy shifts or transition amplitudes, the present approach probes the corresponding interaction through its mechanical force signature in a macroscopic solid-state system.

We show that the combination of a vector force sensor and suitably designed spin configurations allows the pure $e$-$e$ contribution to be isolated from competing mixed $e$-\{$e,N$\} channels. In this light, the commonly adopted assumption that the $e$-$e$ contribution can be neglected \cite{kim_experimental_2019,su_search_2021} should be revisited \cite{yan_new_2013}. More accurately, existing fifth-force experiments probing $V_{12\pm13}$ \cite{heckel_preferred-frame_2008,dzuba_probing_2017,kim_experimental_2019,jiao_experimental_2021,liang_new_2023,wu_searching_2022,clayburn_using_2023,heng_search_2025,tian_constraints_2025} should be interpreted as constraining a combined $e$-\{$e,N$\} interaction rather than a purely $e$-$N$ one. Setting the $e$-$e$ interaction to zero and then deriving bounds on the $e$-$N$ channel is not model independent, since the $e$-$e$ and $e$-$N$ AV interactions may carry opposite signs and partially cancel in the total signal. Conversely, a sufficiently strong direct bound on $g_A^e g_V^e$, relative to $g_A^e g_V^N$, would justify neglecting $e$-$e$ couplings and allow robust bounds on $g_A^e g_V^N$ to be extracted from $V_{12\pm13}$-based searches. Further study of the pure $e$-$e$ contribution is therefore well motivated.

\mysection{Conclusion}

We developed a levitated ferromagnetic vector force sensor as a new tool for probing spin-dependent exotic interactions. By providing experimental access to their directional signatures, levitated magnets allow selective discrimination between interaction channels within the same coupling class \cite{cong_spin-dependent_2025}. As a representative example, we studied AV interactions mediated by a spin-1 boson, such as a $Z’$, in a system of two levitated ferromagnets. Using setup based on an already demonstrated experiment, we isolated the pure electron–electron spin-dependent contribution and derived projected constraints in a previously unexplored short-range regime. Future improvements can enhance the sensitivity by up to several orders of magnitude and extend the present vector-force-sensing strategy beyond AV interactions to other classes of spin-dependent exotic interactions.

\acknowledgments
We would like to thank Francis Headley and Saarik Kalia for enriching discussions.
DA has been supported by ERC grant ERC-2024-SYG 101167211 by the European Union. LC has been supported by the Cluster of Excellence “Precision Physics,
Fundamental Interactions, and Structure of Matter” (PRISMA++ EXC 2118/2) funded
by the German Research Foundation (DFG) within the German Excellence Strategy
(Project ID 390831469) and by the COST Action within the project COSMIC WISPers (Grant No. CA21106). TMF and HU have been supported by the EU Horizon Europe EIC Pathfinder project QuCoM (101046973) under the Innovate UK Horizon Europe Guarantee fund (10032223), by the UK funding agency EPSRC (EP/X009491/1), and by the Leverhulme Trust (RPG-2022-57).
Views and opinions expressed are however those of the author(s) only and do not necessarily reflect those of the European Union, European Research Council Executive Agency, or other awarding body. Neither the European Union nor the granting authority can be held responsible for them.

\acknowledgments

\bibliography{MyLibrary}

\clearpage

\appendix

\onecolumngrid

\begin{center}
\large
\textbf{
    \textit{Supplemental Material:} \\ A Levitated-Magnet Vector Force Sensor for Spin-Dependent Exotic-Interactions
    }
    
\vspace{1.75ex}

Dorian W.~P.~Amaral, Lei Cong, Tim M.~Fuchs, Hendrik Ulbricht, and Dmitry Budker

\vspace{2ex}

\end{center}

\twocolumngrid

\section{A.~Forces from Potentials}
Consider a coordinate system in which the sensing magnet of mass $m_1$ is placed at position vector $\bv{r}_1$ and the driver is placed at position vector $\bv{r}_2$. Their relative displacements and velocities are then given by $\bv{r} \equiv \bv{r}_1 - \bv{r}_2$ and $\bv{v} \equiv \dot{\bv{r}}$, respectively. The driver provides the sensor with a potential energy that can be dependent on both of these quantities, $V(\bv{r}, \bv{v})$, such that the Lagrangian describing the dynamics of the sensing magnet is given by
\begin{equation}
    L = \frac{1}{2} m_1|\dot{\bv{r}}_1|^2 - V(\bv{r}, \bv{v})\,.
\end{equation}
The Euler-Lagrange equation for the $i^\mrm{th}$ coordinate of the magnet is then
\begin{equation}
    \pv{L}{r_1^i} - \dv{}{t}\pv{L}{\dot{r}_1^i} = \pv{L}{r^i} - \dv{}{t}\pv{L}{\dot{r}^i} = 0\,,
\end{equation}
where we have applied the chain rule and used the fact that $\pvl{r^i}{r_1^j} = \delta^{ij}$ and $\pvl{\dot{r}^i}{\dot{r}_1^j} = \delta^{ij}$. Then the generalized force experienced by the magnet in vector form is given by
\begin{equation}
\label{eq:force_euler_lagrange}
    \bv{F}(\bv{r}, \bv{v}) = - \nabla_\bv{r} V(\bv{r}, \bv{v}) + \dv{}{t} [\nabla_\bv{v}V(\bv{r}, \bv{v})]\,.
\end{equation}

The potentials themselves contain operations involving the Pauli matrices $\bs{\sigma}$ of the interacting fermions. Since the spin and spatial degrees of freedom are independent, the Pauli matrix $\bs{\sigma}_i$ for fermion $i$ commutes with all spatial operators. To compute the classical force, we take the expectation value of the resulting force operator with the full two-body spin state $|\chi \rangle \equiv |\chi_1\rangle \otimes |\chi_2\rangle$. Since this is a product state, we may make the replacements $\bs{\sigma}_i \rightarrow\langle \bs{\sigma}_i \rangle \equiv \langle\chi_i| \bs{\sigma}_i |\chi_i\rangle = \hbv{s}_i$, where $\hbv{s}_i$ is a unit vector pointing along the spin-polarization axis of fermion $i$. We therefore write our force expressions in terms of the vectors $\hbv{s}_i$.

\subsection{B.~Force Expressions}
The spin-dependent exotic potentials have been systematically organized in Ref.~\cite{cong_spin-dependent_2025}, where the 16 potentials of Dobrescu and Mocioiu \cite{dobrescu_spin-dependent_2006} are mapped onto nine underlying coupling classes. In the present work, we focus on one such class, namely the axial-vector--vector coupling, which contains the effective potentials $V_{11}$ and $V_{12\pm13}$.
Beginning from the potentials given in \cref{eq:potential}, we can derive the force expressions for a general configuration via \cref{eq:force_euler_lagrange}. The macroscopic force felt by the sensing magnet for $V_{11}|_{AV} + V_{11}|_{VA}$ is then
\begin{align}
    \bv{F}_{11}(\bv{r}) &= g_A^e g_V^e \mcal{N}_\sigma^2\frac{\hbar^2}{4\pi m_e}e^{-r/\lambda} \nonumber\\
    &\times\bigg\{\left(\frac{3}{r^3} + \frac{3}{\lambda r^2} +\frac{1}{\lambda^2 r}\right)\big[(\hbv{s}_e \times \hbv{s}_e') \cdot \hbv{r}\big]\hbv{r} \\
    &-\left(\frac{1}{r^3} - \frac{1}{\lambda r^2}\right)(\hbv{s}_e \times \hbv{s}_e')\bigg\}\,,\nonumber
\end{align}
and for $V_{12+13}|_{AV} + V_{12-13}|_{AV}$ we find
\begin{align}
    \bv{F}_{12\pm13}(\bv{r},\bv{v})
    &= - g_A^e g_V^{e+N} \mcal{N}_\sigma \mcal{N}_f\frac{\hbar}{4\pi}e^{-r/\lambda} \\
    &\times\left(\frac{1}{r^2} + \frac{1}{\lambda r}\right) \,\big\{\bv{v} \times \big[(\hbv{s}_{e} - \hbv{s}_{e}') \times \hbv{r}\big]\big\}
\,. \nonumber
\end{align}
For the configuration we construct in our study, whereby $\hbv{r} = \hbv{x}$, $\hbv{s}_e = \hbv{y}$, $\hbv{s}_e' = \hbv{x}$ and $\bv{v} = v(t)\hbv{x}$, these expressions reduce to the forms we have in \cref{eq:forces}.

The above derived forces assume that the magnets can be treated as point-like particles. This is valid for force ranges larger than their radii, $\lambda \gg R$. To go beyond this approximation and model finite-size effects, we would proceed by assuming spin and nucleon number densities for each magnet and use the relevant Green's function for each of the potentials when solving for the forces.

To estimate the number of polarized electron spins in our magnets, we consider their remnant magnetic fields $\bv{B}_r$. For a $\mrm{Nd_2 Fe_{14} B}$ magnet, the total magnetic moment is given by
\begin{equation}
\begin{split}
    \bs{\mu}_s &= \frac{\bv{B}_r V_s}{\mu_0} \\
    &\simeq 5.2\times10^{-5}\,\mrm{J\,T^{-1}} \left(\frac{\bv{B}_r}{1.4\,\mrm{T}}\right)\left(\frac{m_s}{0.356\,\mrm{mg}}\right)\,,
\end{split}
\end{equation}
where we have fiducialized to the remnant field of a $\mrm{Nd_2 Fe_{14} B}$ magnet and converted to the magnet mass $m_s$ from its volume $V_s$ using its density, $\rho_s \approx 7600\,\mrm{kg\,m^{-3}}$. Then, using the magnetic moment of the electron $\mu_e \approx 9.3 \times 10^{-25}\,\mrm{J\,T^{-1}}$, we find
\begin{equation}
    N_\sigma = \frac{\mu_s}{\mu_e} \simeq 6\times 10^{18}\left(\frac{m_s}{0.356\,\mrm{mg}}\right)\,.
\end{equation}

\section{C.~Linearizing the Equations of Motion}
Consider the equation of motion for the levitated magnet, which follows that of a driven, damped harmonic oscillator for motions smaller than the extent of the trapping potential. The sensing magnet of mass $m_1$ is placed such that its equilibrium position is at the origin, with small displacements from it following the position vector $\bv{r}_1$. The driving magnet is placed with its equilibrium position at $\bv{r}_2$ and driven at a velocity of $\bv{v}_2$. As we argued above, the sensing magnet is driven by a force $\bv{F}$ that depends on the relative position and velocity vectors, $\bv{r} \equiv \bv{r}_1 - \bv{r}_2$ and $\bv{v} = \dot{\bv{r}}$. We may then write the equation of motion for the sensing magnet as
\begin{equation}
\ddot{\bv{r}}_1(t) + \Gamma\dot{\bv{r}}_1(t) + \Omega\bv{r}_1(t) = \frac{\bv{F}\textbf{(}\bv{r}(t), \bv{v}(t)\textbf{)}}{m_1}\,,
\label{eq:sensor_eom}
\end{equation}
where $\Omega \equiv \mrm{diag}(\omega_x^2, \omega_y^2, \omega_z^2)$ is a diagonal matrix containing the angular frequencies for each mode and $\Gamma \equiv \mrm{diag}(\gamma_x, \gamma_y, \gamma_z)$ is a diagonal damping rate matrix. We take  $ \gamma_i \equiv \omega_i/Q_i$, with $Q_i$ the quality factor of the $i^\mrm{th}$ mode. 

Now let us expand the force felt by the sensing magnet around its equilibrium position for small displacements $\bv{r}_1$. To linear order, we may write the force as
\begin{equation}
\begin{split}
&\bv{F}\textbf{(}\bv{r}, \bv{v}\textbf{)} \simeq \bv{F}\textbf{(}-\bv{r}_2, -\bv{v}_2\textbf{)} \\&+ \pv{\bv{F}}{\bv{r}}\bigg|_{\textbf{(}-\bv{r}_2, -\bv{v}_2\textbf{)}} \bv{r}_1(t)
+  \pv{\bv{F}}{\bv{v}}\bigg|_{\textbf{(}-\bv{r}_2, -\bv{v}_2\textbf{)}}\dot{\bv{r}}_1(t)\,,
\end{split}
\end{equation}
where $(\partial \bv{F}/\partial\bv{r})_{ij} \equiv \partial F_i/\partial r_j$ and $(\partial \bv{F}/\partial\bv{v})_{ij} \equiv \partial F_i/\partial v_j$. Defining the matrices
\begin{equation*}
\Omega'(t) \equiv \frac{1}{m_1}\pv{\bv{F}}{\bv{r}}\bigg|_{\textbf{(}-\bv{r}_2, -\bv{v}_2\textbf{)}} ~~ \text{and} ~~ \Gamma'(t) \equiv \frac{1}{m_1}\pv{\bv{F}}{\bv{v}}\bigg|_{\textbf{(}-\bv{r}_2, -\bv{v}_2\textbf{)}}\,,
\end{equation*}
the equation of motion may be rewritten as
\begin{equation}
\ddot{\bv{r}}_1(t) + \tilde{\Gamma}(t)\dot{\bv{r}}_1(t) + \tilde{\Omega}(t)\bv{r}_1(t) = \frac{\bv{F}\textbf{(}-\bv{r}_2(t), -\bv{v}_2(t)\textbf{)}}{m_1}\,,
\end{equation}
where we now have the dynamical coefficients $\tilde{\Gamma}(t) \equiv \Gamma - \Gamma'(t)$ and $\tilde{\Omega}(t) \equiv \Omega - \Omega'(t)$, introducing backreaction. 

This backreaction modifies the equation of motion for the sensing magnet from that of a simple, damped harmonic oscillator. However, they can be ignored as long as the below conditions are met:
\begin{equation}
    \begin{split}
    \left|\sum_j \Gamma'_{ij}(t) \bv{r}_{1,j}(t)\right| &\ll \gamma_i |\bv{r}_1(t)|\,,\\
    \left|\sum_j \Omega'_{ij}(t) \bv{r}_{1,j}(t)\right| &\ll \omega_i^2 |\bv{r}_1(t)|\,.
    \end{split}
\end{equation}
Since in our analysis we consider motions that preferentially excite one mode at a time, we may simplify these expressions to requiring that
\begin{equation}
    \begin{split}
    \left|\Gamma'_{ii}(t)\right|/\gamma_i &\ll 1\,,\\
    \left|\Omega'_{ii}(t)\right| / \omega_i^2 &\ll 1 \,,
    \end{split}
\end{equation}
where no sum is implied. If these conditions are met, then the equation of motion for the sensing magnet follows that of driven oscillator with the force evaluated using only the trajectory of the driving magnet, allowing us to ignore the motion of the sensing magnet entirely. We verify that this condition is satisfied for all potentials considered, justifying the approximation throughout our analysis.

\section{D.~Complete Sensitivity Analysis}

\begin{figure}
    \centering
    \includegraphics{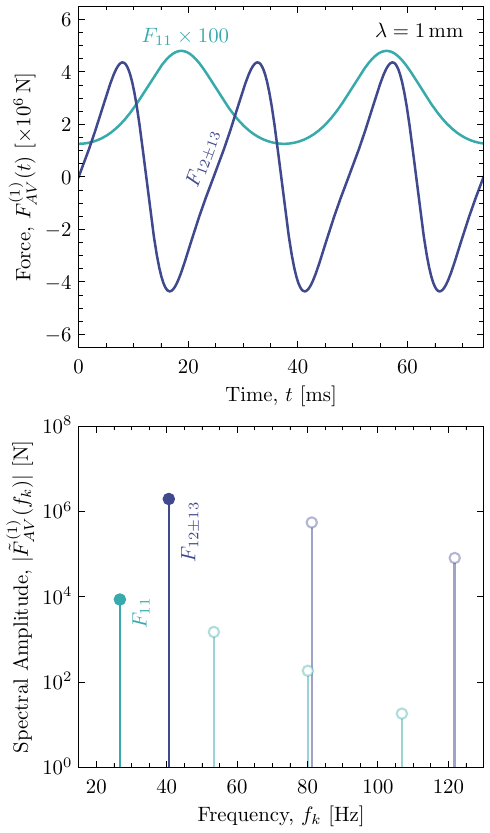}
    \caption{The exotic force signals we consider in \cref{eq:forces} sourced due to the sinusoidal motion of the driver. We show this  for unit couplings and for a force range $\lambda = 1\,\mathrm{mm}$ in the time and Fourier domains. \textbf{Top}: The signal in the time domain $F_{AV}^{(1)}(t)$. \textbf{Bottom:} The magnitude of the  signal in the Fourier domain $|\tilde{F}^{(1)}_{AV}(f_k)|$, defined as the spectral force amplitudes and computed via \cref{eq:fourier_average}. The solid peaks highlight the signals that are resonantly enhanced by our sensor. The faint peaks show the signals present at higher frequencies, which are unimportant for our study.}
    \label{fig:spec_amp}
\end{figure}

We define the spectral amplitude of the force $F(t)$ to be its time normalized physical Fourier transform. For a given choice of parameter values $\bs{\theta}$, where $\bs{\theta} \equiv (g_A^e g_V^{e/e+N}, \lambda)^\intercal$, we have
\begin{equation}
    \tilde{F}_{AV}(\omega_k, \bs{\theta}) \equiv \frac{\Delta t}{T_\mrm{num}} \sum_{n=0}^{N - 1}F_{AV}(t_n, \bs{\theta}) e^{-i \omega_k t_n}\,,
    \label{eq:fourier_average}
\end{equation}
where the choice of coupling is $g_A^e g_V^e$ for $F_{AV} = F_{11}$ and $g_A^e g_V^{e+N}$ for $F_{AV} = F_{12\pm 13}$. Here, $\omega_k$ is the angular frequency in the $k^\mrm{th}$ frequency bin, $T_\mrm{num}$ is the total time for which we numerically compute the signal in the time domain, $N$ is the number of points sampled in this domain, $\Delta t \equiv T_\mrm{num} / N$ is the sampling interval, and $t_n \equiv n \Delta t$. We show the spectral amplitudes $\tilde{F}_{11}(\omega_k)$ and  $\tilde{F}_{12\pm13}(\omega_k)$ for unit couplings and a large force range of $\lambda = 1\,\mrm{m} \gg d$ in \cref{fig:spec_amp}. We observe peaks of decaying amplitudes at integer multiples of the driving frequency. For our analysis, we consider only the peak occuring at the fundamental frequency, to which we tune our sensing magnet, since this carries the strongest signal. Their numerical values in the limit that $\lambda \rightarrow \infty$ are $|\tilde{F}_{11}| \simeq g_A^e g_V^e (6 \times 10^4\,\mrm{N})$ and $|\tilde{F}_{12\pm13}| \simeq g_A^e g_V^{e+N} (3 \times 10^7\,\mrm{N})$

We employ a matched filtering analysis to compute our sensitivities. To account for systematic uncertainties in our experiment, we include a pull parameter $\xi$ as a nuisance parameter in our likelihood, acting to linearly scale the signal template.  Given a measured timeseries $d(t)$ and a signal template $F_{AV}(t)$, our log-likelihood is then
\begin{equation}
\begin{split}
    \ln\mcal{L}(\bs{\theta}, \xi;d) &= (1 + \xi)\langle d, \tilde{F}_{AV} \rangle \\
    &-\frac{(1 + \xi)^2}{2} \langle \tilde{F}_{AV}, \tilde{F}_{AV} \rangle
    - \frac{\xi^2}{2 \sigma_\xi^2}\,,
\end{split}
\end{equation}
where the matched filter inner product for the arbitrary templates $a(t)$ and $b(t)$ is defined as
\begin{equation}
    \langle a, b\rangle \equiv \frac{4}{T_\mrm{obs}} \Re \left\{\sum_{k} \frac{\tilde{a}^*(\omega_k) \tilde{b}(\omega_k)}{S_{FF}(\omega_k)}\right\}\,,
    \label{eq:inner_prod}
\end{equation}
with $T_\mrm{obs}$ the total observation time since this governs the frequency resolution of the final signal. Ultimately, since we focus only on the peak on resonance, the above sum collapses to the single term at $\omega_k = \omega_d$. 

To go further analytically, we first define the `unit-coupling' template $F_{AV}^{(1)}(\lambda)$  as that which we compute via \cref{eq:fourier_average} by setting $g_A^e g_V^{e/e+N} = 1$ for a given choice of force range $\lambda$. This allows us to write $\langle \cdot, F_{AV}\rangle = g_A^e g_V^{e/e+N}\langle \cdot, F_{AV}^{(1)}\rangle$ using the linearity of the inner product. Next, we assume an Asimov dataset, whereby we replace the data $d$ with its expectation value under the true hypothesis~\cite{cowan_asymptotic_2011}. Since we derive sensitivity projections in this study, the true hypothesis is that no signal is present. We thus set $d = 0$, as no gravitational signal will be present in the directions that either one of the exotic forces act and the expectation value of the signal due to random noise is zero. Our likelihood therefore reduces to
\begin{equation}
\begin{split}
    \ln \mcal{L}(\bs{\theta}, \xi) = &-\frac{(1 + \xi)^2}{2} (g_A^e g_V^{e+N})^2 \langle {F}_{AV}^{(1)}, {F}_{AV}^{(1)}\rangle \\
    &- \frac{\xi^2}{2 \sigma_\xi^2}\,.
\end{split}
\end{equation}

We use this likelihood to perform our inference via the one-sided test statistic
\begin{equation}
    q \equiv -2 \ln\left[\frac{\mcal{L}(\bs{\theta}, \hat{\xi})}{\mcal{L}(\hat{\hat{\bs{\theta}}}, \hat{\hat{\xi}})}\right]\,,
\end{equation}
where double-hatted variables indicate the maximum-likelihood-estimators of the unconstrained likelihood, and the single-hatted variable signifies the maximum-likelihood-estimator of the likelihood conditioned on a particular value of $\bs{\theta}$, forming the profiled likelihood. Larger values of $q$ signify larger disagreements with our data. Since we assume an Asimov data set with $d = 0$, the unconstrained likelihood is simply maximized for $g_A^e g_V^{e/e+N} = 0$, yielding $\mcal{L}(\hat{\hat{\bs{\theta}}}, \hat{\hat{\xi}}) = 0$. Solving for $\hat{\xi}$, the test statistic can then easily be shown to be
\begin{equation}
    q = \frac{(g_A^e g_V^{e/e+N})^2\langle {F}_{AV}^{(1)}, {F}_{AV}^{(1)}\rangle}{1 + \sigma_\xi^2(g_A^e g_V^{e/e+N})^2\langle {F}_{AV}^{(1)}, {F}_{AV}^{(1)}\rangle}\,.
\end{equation}

We may solve this expression for a given choice of $\lambda$ to find an analytical expression for our coupling sensitivities. Using the definition of the inner product in \cref{eq:inner_prod}, we find:
\begin{equation}
\begin{split}
    \left\lvert g_A^e g_V^{e/e+N}(\lambda)\right\rvert &\gtrsim  \frac{\mcal{T}(T_\mrm{obs}/\tau)}{2|\tilde{F}_{AV}^{(1)}(\lambda)|}\\ &\times\sqrt{\left(\frac{q_\mrm{}}{1 - q\sigma_\xi^2}\right)\frac{S_{FF}^\mrm{th}(\omega_d)}{T_\mrm{obs}}}\,,
\end{split}
\label{eq:sensitivity_full}
\end{equation}
where we have re-introduced the notation $F_{AV}^{(1)}$ to symbolize one of our two exotic forces. To re-implement the scaling of the signal with observation time, we have introduced the dimensionless scaling factor $\mcal{T}(T_\mrm{obs}/\tau)$. This function captures how the signal PSD grows with time due to the gradual ring-up of the resonator. We derive this function in the next section; however, its parametric behavior is such that $\mcal{T} \rightarrow 1$ for $T_\mrm{obs}/\tau \gg 1$ and $\mcal{T} \propto (T_\mrm{obs} / \tau)^2$ for $T_\mrm{obs}/\tau \ll 1$, when the resonator is still ringing up. 

Finally, we must choose the value of $q$. To compute our sensitivity at a chosen confidence level, we must be able reject the null hypothesis that a signal exists over background-only data; this sets a lower threshold on the value of $q$. Since the data themselves are random variables, $q$ is also a random variable with some probability density function conditioned on the parameter space point, $f(q|\bs{\theta})$. The limiting threshold for $q$, $q_\mrm{lim}$, is dictated by the $p$-value
\begin{equation}
    p \equiv \int_{q_\mrm{lim}}^\infty f(q|\bs{\theta}) \dd q\,,
\end{equation}
defining the probability of finding a value of $q$ at least as extreme as the one observed. Following the Asimov result of Ref.~\cite{cowan_asymptotic_2011}, $q$ follows a half-$\chi^2$ distribution $f(q|\bs{\theta}) = (1/2)\delta(0) + (1/2) \chi^2_1(q)$, where $\chi^2_1$ is a chi-square function with one degree of freedom. This follows from Chernoff's theorem, which is relevant when the true parameter value lies at the edge of a parameter space \cite{Chernoff_On_1954}. We may then solve the above integral equation analytically, yielding  $q_\mrm{lim} \approx 2.71$ for a $95\%$ C.\,L. limit, corresponding to a $p$-value of $p=0.05$. This gives us all the necessary information to draw our sensitivity curves.

For the value $q = q_\mrm{lim}$, we are set to be statistics dominated for any reasonable assumption on the systematic uncertainty $\sigma_\xi$. From the parenthetical expression in \cref{eq:sensitivity_full}, we see that $(1-q\sigma_\xi^2)^{-1/2} \simeq 1$ for any $\sigma_\xi \lesssim 50\%$, such that we may safely neglect this correction factor. This yields the ultimate expression we write in \cref{eq:sensitivity}.

\subsection{E.~Accounting for Finite Ring-Up Time}

To account for the ring-up time of our levitated particle for finite observation time $T_\mrm{obs}$, we consider the dynamics of a driven, damped simple harmonic oscillator with resonance frequency $\omega_0$ that is driven by a monochromatic force on resonance. Given a driving force of the form $F_d(\omega_0) \equiv F_0 \cos(\omega_0 t)$, where $F_0$ is the force amplitude, the displacement of the oscillator about its equilibrium position with time, $x(t)$, can be solved from the equation of motion given in \cref{eq:sensor_eom}. 
Assuming the initial values of $x(0) = 0$ and $\dot{x}(0) = 0$ and taking the limit that  $Q \gg 1$, the displacement follows
\begin{equation}
    x(t) = \frac{Q F_0}{m_s \omega_0^2} \sin(\omega_0 t) (1 - e^{-t/\tau})\,,
\end{equation}
where the characteristic damping time is given by $\tau \equiv 2 Q / \omega_0$. 

Taking the finite-time Fourier transform of this expression $\hat{x}_T(\omega)$, we then find the signal periodogram $\mcal{P}$ to be
\begin{equation}
    \mcal{P}(\omega=\omega_0) = \frac{2|\hat{x}_{T}(\omega_{0})|^2}{T_{\mathrm{obs}}} = \frac{A^2 T_\mrm{obs}}{2} \mcal{T}(T_\mrm{obs})\,,
    \label{eq:periodogram_res}
\end{equation}
where the amplitude is given by $A \equiv Q F_0 / (m_s \omega_0^2)$ and the dimensionless scaling factor
\begin{equation}
    \mcal{T}(x) \equiv \left(1 - \frac{1 - e^{-x}}{x}\right)^{2}
    \label{eq:growth_factor}
\end{equation}
is the growth factor we introduced in \cref{eq:sensitivity}.
In the limiting regime that the observation time is much shorter than the damping time, the resonator is  still ringing up and its displacement amplitude grows linearly with time. Conversely, for much larger observation times, the resonance peak grows only by virtue of the improving frequency resolution with increasing observation time:
\begin{align}
\qquad~~\mathcal{P}(\omega=\omega_{0}) \simeq \frac{A^{2}T_{\mathrm{obs}}^{3}}{8 \tau^{2}} &&(T_{\mathrm{obs}} \ll \tau)\,, \nonumber \\
\mathcal{P}(\omega=\omega_{0}) \simeq \frac{A^{2}T_{\mathrm{obs}}}{2} &&(T_{\mathrm{obs}} \gg \tau)\,.
\label{eq:peak_scaling}
\end{align}

\begin{figure}
    \centering
    \includegraphics{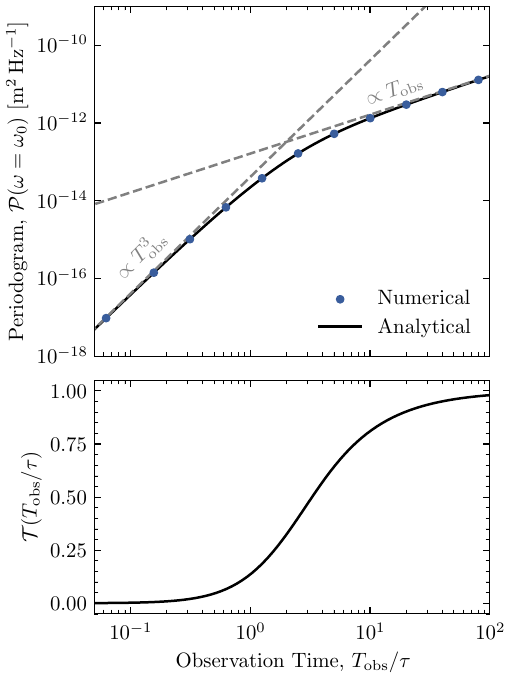}
    \caption{\textbf{Top:} The periodogram peak on resonance $\mcal{P}(\omega = \omega_0)$ with increasing observation time $T_\mrm{obs}$ in units of the characteristic damping time $\tau \equiv 2 Q / \omega_0$. The analytical result follows from \cref{eq:periodogram_res}, where the dependence on observation time is captured by the growth factor. Also shown are the power-law relations in the limiting regimes that $T_\mrm{obs}/\tau \ll 1$ and $T_\mrm{obs} / \tau \gg 1$. \textbf{Bottom:} The growth factor $\mathcal{T}(T_\mrm{obs}/\tau)$, as given in  \cref{eq:growth_factor}. For $T_\mrm{obs}/\tau \ll 1$, the resonator has fully rung up, and the increase in the signal periodogram is purely due to the improving frequency resolution with increasing observation time.}
    \label{fig:ring_up}
\end{figure}

We verify this result by numerically solving the ordinary differential equation given by the equation of motion and computing the periodogram for varying observation times. We take $F_0 = 10\,\mrm{N}$, $m_s = 1\,\mrm{kg}$, $f_0 = 10\,\mrm{Hz}$ with $\omega_0 \equiv 2 \pi f_0$, and $Q = 500$. We show our results in \cref{fig:ring_up}, where we find excellent agreement between the numerical and analytical results.  In our sensitivity analysis, the prefactor $A^2 / 2$ is replaced with the numerically computed Fourier component given by \cref{eq:fourier_average}.

\section{F.~Future Configuration}
To improve the sensitivity of the experiment in a future configuration, we reduce the sensor to two parameters that uniquely determine all other quantities that impact our sensitivities under certain assumptions. These quantities are the particle radius $R_s$ and the trap semi-major axis $a$.

The first assumption is that the magnetic particles are spherical, which means that we can determine the mass, charge and magnetic moment of the particle directly from the radius through the volume, density and magnetization. In this work, we assume Nd$_2$Fe$_{14}$B magnets, magnetized to a remanent magnetization of $B_{rem}=\SI{}{T}$.
For such a spherical magnetic dipole, we can also determine the trap frequency from $R_s$ and $a$ by following the analytical method developed by Ref.~\cite{headley_magnetic_2025}, which uses an infinite series of induced dipoles to calculate the frequency from the magnetic potential under perfect Meissner repulsion. For a particle placed between two infinite superconducting planes at a distance the $d$ from the particle, the translational mode frequencies are found as 
    \begin{equation}
    \begin{split}
        \omega_{q_\parallel} &= \sqrt{\frac{93}{256}\frac{\zeta(5)}{\pi}\frac{\mu_0\mu^2}{m_s d^5}}\quad \text{and} \quad \omega_{q_\perp} = \sqrt{2}\omega_{q_\parallel}\,.
    \end{split}
    \label{eq:mode_freqs}
    \end{equation}
Here, $q_\parallel$ is the translational mode along which the magnetic moment is pointed and $q_\perp$ is the mode perpendicular to it. In these expressions, $d$ takes the role of either the semi-major or semi-minor axis depending on the orientation of the trap and which mode is being calculated. The quantities $\mu$ and $m_s$ are the magnetic moment and mass of the particle, respectively, which are determined by $R_s$. The Riemann zeta function $\zeta(n)$ arises from the converging infinite series of dipoles.

The only quantity of import left to express as a function of $R_s$ and $a$ is the quality factor $Q$ of the resonator. For this, we make the final assumption that the damping of the system is determined predominantly by eddy-current damping induced by the motion of the magnetic particle. Since eddy-current damping scales as $\omega^{-2}$, we can scale the quality factors reported in the previous work of Ref.~\cite{fuchs_2024_gravity}. As the frequency is determined by $R_s$ and $a_s$, this scaling allows us to express the quality factor as function of these two quantities.

Even if the system were not eddy-current limited, the $\omega^{-2}$ scaling provides a good lower limit to the $Q$ factors of the system as long as they are only scaled in such a way that the quality factor is reduced. We do not extrapolate towards higher quality factors/lower damping, as the scaling cannot be guaranteed to hold in this regime. For the future setup, we assume that further improvements to trap and particle preparation are expected to increase the $Q$-factor by another order of magnitude over the current state of the art.

Combining these results, we can approximate how our sensitivities scale with the particle radii and the trap dimensions. First, the thermal noise of power spectral density, defined in \cref{eq:thermal_noise}, scales as
\begin{equation}
\begin{split}
    S_{FF}^\mrm{th} &\sim 10^{-32}\,\mrm{N^2\,Hz^{-1}} \\
    &\times\left[\left(\frac{R_s}{236\,\mrm{\mu m}}\right)\left(\frac{2.25\,\mrm{mm}}{a}\right)\right]^{15/2},
\end{split}
\end{equation}
where we have fiducialized to the configuration of the presently realizable experiment.
This indicates that the noise will worsen with increasing particle size but improve with larger traps. This strong scaling with the trap dimension, which is itself due to our assumption on how the quality factor scales, is the reason why larger traps tend to be better for our sensitivities.

For simplicity, we assume that we are in the fully rung-up regime ($T_\mrm{obs} / \tau \gg 1$) and in the long force range limit ($\lambda \rightarrow \infty$). The signal PSD then scales as $\mcal{P} \simeq \Delta F_{AV}^2 T_\mrm{obs} / 2$, where the force modulation is given by
\begin{equation}
\begin{split}
    \Delta F_{AV} &= \mcal{F}_{AV} \left[\frac{1}{(\langle r \rangle -\mcal{A})^n} - \frac{1}{(\langle r \rangle +\mcal{A})^n}\right]\\
    &\simeq \mcal{F}_{AV} \left(\frac{n \epsilon}{2^n a^n}\right) \propto \mcal{F}_{AV} a^{-n}\,.
\end{split}
\end{equation}
Here, we have taken the average separation to be $\langle r \rangle = a + b + \delta \simeq 2 a$ since $a \sim b > \delta$, and we have set the amplitude of the driver to be a fraction of the semi-major axis, $\mcal{A} = \epsilon a$, with $\epsilon = 0.2$ in our work. The quantity $\mcal{F}_{AV}$ is the relevant prefactor for either $F_{11}$ or $F_{12\pm13}$ given by \cref{eq:forces}, and $n$ is the relevant distance power scaling---$n=3$ for $F_{11}$ and $n = 2$ for $F_{12\pm13}$. In both cases, $\mcal{F}_{AV}$ scales with the sizes of both the driver and the sensor since both are proportional to summed fermion numbers; therefore, $\mcal{F}_{AV} \propto (R^3)^2 $. However, for $F_{12\pm13}$, there is an additional scaling by the velocity $v \sim \mcal{A} \omega_d \sim \epsilon a \omega_i \propto (R/a)^{3/2}$, where we have set $\omega_d = \omega_i$ to achieve resonance of the relevant mode $i$ and used the frequency relation in \cref{eq:mode_freqs} with $d \sim a$. Combining all of these scalings, we then have that
\begin{equation}
\Delta F_{AV} \propto
    \begin{cases}
        g_A^e g_V^e R^6 a^{-3} & \text{for $F_{11}$}\,,\\
        g_A^e g_V^{e+N} R^{15/2} a^{-7/2} & \text{for $F_{12\pm13}$}\,.
    \end{cases}
\end{equation}

Finally, using these expressions and the fact that the SNR is approximately given by $\mrm{SNR} \sim \mcal{P}/S_{FF}^\mrm{th}\propto \Delta F^2_\mrm{AV} / S_{FF}^\mrm{th}$, we arrive at the scalings for our sensitivities:
\begin{align}
        \label{eq:g_scalings}
        &\!\!\!\left\lvert g_A^e g_V^e \right\rvert \simeq 6 \times 10^{-24}\,\left(\frac{236\,\mrm{\mu m}}{R_s}\right)^{9/4}\left(\frac{2.25\,\mrm{mm}}{a}\right)^{3/4}\,,\\
        &\!\!\! \left\lvert g_A^e g_V^{e+N} \right\rvert \simeq 1 \times 10^{-26}\,\left(\frac{236\,\mrm{\mu m}}{R_s}\right)^{15/4}\left(\frac{2.25\,\mrm{mm}}{a}\right)^{1/4}\,,\nonumber
\end{align}
where we have set the coupling values to those of our sensitivities in \cref{fig:sensitivity} in the large $\lambda$ limit for the relevant force. These expressions provide us an approximate scaling for our sensitivities since they assume that the resonator has completely rung up. However, for high mode frequencies, the quality factors increase, leading to this assumption breaking down when the characteristic damping time is such that $\tau_i \equiv 2 Q_i / \omega_i \propto 1/\omega_i^3 \gg T_\mrm{obs}$, as shown in \cref{eq:peak_scaling}. Indeed, for the $z$-mode in the present setup, the resonator is still ringing up. The scalings for arbitrary observation times is complex; nevertheless, they can easily be retrieved by multiplying \cref{eq:g_scalings} by the peak growth factor $\mcal{T}(T_\mrm{obs}/\tau)$ given in \cref{eq:growth_factor}. The result is that larger particles always act to improve our sensitivities, but it is more advantageous to have smaller traps until the resonator is able to completely ring up. In this regime, larger traps are then more beneficial due to the decrease in the thermal force noise. 

\end{document}